\documentclass[pdflatex,sn-mathphys-num]{sn-jnl}

\usepackage{graphicx}%
\usepackage{multirow}%
\usepackage{amsmath,amssymb,amsfonts}%
\usepackage{amsthm}%
\usepackage[title]{appendix}%
\usepackage{xcolor}%
\usepackage{textcomp}%
\usepackage{manyfoot}%
\usepackage{booktabs}%
\usepackage{algorithm}%
\usepackage{algorithmicx}%
\usepackage{algpseudocode}%
\usepackage{listings}%
\usepackage{lmodern}
\usepackage{bm}

\begin{document}

\title[Title]{Compact Model of Linear Passive Integrated Photonics Device for Photon Design Automation[Work In Progress]}


\author*[1]{\fnm{Zijian} \sur{Zhang}}\email{hectorzhang4253@gmail.com}

\affil*[1]{\orgname{University of Electronic Science and Technology of China}, 
\orgaddress{\city{Chengdu}, \postcode{611731}, \country{China}}}


\abstract{As integrated photonic systems grow in scale and complexity, Photonic Design Automation (PDA) tools and Process Design Kits (PDKs) have become increasingly important for layout and simulation. However, fixed PDKs often fail to meet the rising demand for customization, compelling designers to spend significant time on geometry optimization using FDTD, EME, and BPM simulations. To address this challenge, we propose a data-driven Eigenmode Propagation Method (DEPM) based on the unitary evolution of optical waveguides, along with a compact model derived from intrinsic waveguide Hamiltonians. The relevant parameters are extracted via complex coupled-mode theory. Once constructed, the compact model enables millisecond-scale simulations that achieve accuracy on par with 3D-FDTD, within the model’s valid scope. Moreover, this method can swiftly evaluate the effects of manufacturing variations on device and system performance, including both random phase errors and polarization-sensitive components. The data-driven EPM thus provides an efficient and flexible solution for future photonic design automation, promising further advancements in integrated photonic technologies.}




\maketitle
\newpage
\section{Introduction}\label{sec1}
The rise of integrated photonics has raised the demand for highly integrated and customized waveguide devices in large-scale optoelectronic chips\cite{shekhar_roadmapping_2024}. Compared with conventional electronic chips, integrated photonic devices offer significant advantages in parallel processing and low-latency interconnects, and have been widely used in deep learning\cite{shen_deep_2017}, quantum secure communication\cite{paraiso_photonic_2021}, optical switching networks\cite{chen_review_2023,qiao_32_2017}, and Lidar\cite{hsu_review_2021} at al.

Waveguide devices are extremely sensitive to geometric variations, especially for high-index-contrast waveguides. Slight deviations in waveguide width or coupling gap can affect the effective index, coupling coefficients, and mode profiles, ultimately causing unpredictable systematic deviations. To achieve “design for manufacturing” (DFM), Electronic-Photonic Design Automation (EPDA) tools have emerged, relying on physical simulations and foundry PDK data to build S-parameter-based compact models for each device, completed the simulation at the function level, and finally automatic cell placement and photonic routing.

In 2018, Bogaerts Wim and colleagues presented five key challenges for EPDA\cite{bogaerts_silicon_2018}: capturing variability effects, circuit and signal representation, building reliable compact models, photonic-electronic co-design, and photonic routing. Over the past seven years, researchers have proposed various solutions, including variation-aware compact model of strip waveguides\cite{james_process_2023}, complex vector fitting for waveguide devices\cite{ullrick_wideband_2023}, phase-change-based photonic device modeling\cite{carrillo_system-level_2021}, and time-domain compact models for thermo-optic phase shifters\cite{coenen_analysis_2022}. These efforts have enabled wafer-scale predictions of operating point drifts in micro-ring resonators (MRRs) and Mach-Zehnder interferometers (MZIs)\cite{bogaerts_predicting_2019}.

Nevertheless, the demand for highly customized waveguide devices and the necessity of accurately solving Maxwell’s equations pose significant computational challenges during iterative design processes. While FDTD, EME, and BPM methods are capable of accurately computing optical fields and eigenmodes\cite{gallagher_eigenmode_2003}, they are resource-intensive and time-consuming. As optimization iterations increase, the computational burden can quickly escalate. In 2022, Ian M. Hammond et al. introduced a deep learning-enhanced FDE and EME toolbox\cite{hammond_deep_2022}. Through machine learning, the EME simulation process is accelerated by 3 times. In 2024, Mehmet Can Oktay et al. introduced a data-driven EME approach\cite{oktay_computationally_2024} that leverages a precomputed scattering matrix database to achieve near–3D-FDTD accuracy within seconds.However, as a "black box" model, their method lacks robust explanations of the underlying physical mechanisms and convergence properties, functioning primarily as a numerical acceleration strategy.

To simultaneously accommodate manufacturing variability assessments, rapid geometry optimization, and compact model extraction, this work adopts a unitarity-based view of light field, introducing a data-driven eigenmode propagation method (DEPM) and a compact model derived from the waveguide intrinsic Hamiltonian. The parameters of the compact model can be obtained by lookup table or fit by parametric method.Through precomputation, DEPM attains computational speed comparable to data-driven EME. By analyzing how the waveguide Hamiltonian shifts with geometric parameters, we can reconstruct the overall evolution from limited data points, and ensure the unitarity of bidirectional propagation. Moreover, DEPM achieves faster convergence compared with traditional EME.

\section{Compact Model Theory}\label{sec2}

For a linear, lossless, and passive photonic waveguide device, its control over optical field transmission can be regarded as a unitary evolution applied to the input field. The corresponding state vector is composed of the normalized complex amplitudes of the waveguide eigenmodes and satisfies a Schrödinger-like equation whose Hamiltonian depends on the waveguide geometry, material properties, and wavelength. A natural idea then follows: if we can extract a set of parameter that is independent of how individual waveguide sections are concatenated—but depends only on the intrinsic waveguide parameters (such as width, height, or coupling spacing), and if this set of parameters allows us to reconstruct the Hamiltonian, we can thereby establish a compact model for describing linear, lossless, and passive waveguide systems.

In 2011, Jianwei Mu and Wei-Ping Huang proposed the theory of complex coupled modes\cite{mu_complex_2011}, which can proposed complex coupled mode theory, which can analytically analyze the coupling behavior of orthogonal eigenmodes (or eigensupermodes) in waveguide systems due to structural changes. The complex coupled mode equation is \ref{eq:CME}.

\begin{equation}
j\frac{d}{dy}\left[ \begin{matrix}
   {{\varphi }^{+}}  \\
   {{\varphi }^{-}}  \\
\end{matrix} \right]
=
\left[ \begin{matrix}
   \mathbf{B}-j{{\mathbf{K}}^{+}} & -j{{\mathbf{K}}^{-}}  \\
   -j{{\mathbf{K}}^{-}} & -\mathbf{B}-j{{\mathbf{K}}^{+}}  \\
\end{matrix} \right]
\left[ \begin{matrix}
   {{\varphi }^{+}}  \\
   {{\varphi }^{-}}  \\
\end{matrix} \right]. \label{eq:CME}
\end{equation}

where
\begin{align}
  K_{mn}^{+} &= \frac{\iint\limits_{xOz}{\left( \frac{\partial {{\mathbf{E}}_{n}}}{\partial y}\times \mathbf{H}_{m}^{\dagger }+\mathbf{E}_{m}^{\dagger }\times \frac{\partial {{\mathbf{H}}_{n}}}{\partial y} \right)}\cdot d{{\mathbf{S}}_{zx}}}
  {\iint\limits_{xOz}{\left( {{\mathbf{E}}_{m}}\times \mathbf{H}_{m}^{\dagger }+\mathbf{E}_{m}^{\dagger }\times {{\mathbf{H}}_{m}} \right)}\cdot d{{\mathbf{S}}_{zx}}} 
  = -{{\left( K_{nm}^{+} \right)}^{\dagger }}, \label{param:CME1}\\
  K_{mn}^{-} &= \frac{\iint\limits_{xOz}{\left( \frac{\partial {{\mathbf{E}}_{n}}}{\partial y}\times \mathbf{H}_{m}^{\dagger }-\mathbf{E}_{m}^{\dagger }\times \frac{\partial {{\mathbf{H}}_{n}}}{\partial y} \right)}\cdot d{{\mathbf{S}}_{zx}}}
  {\iint\limits_{xOz}{\left( {{\mathbf{E}}_{m}}\times \mathbf{H}_{m}^{\dagger }+\mathbf{E}_{m}^{\dagger }\times {{\mathbf{H}}_{m}} \right)}\cdot d{{\mathbf{S}}_{zx}}} 
  = -{{\left( K_{nm}^{-} \right)}^{\dagger }},\label{param:CME2} \\
  {{B}_{mn}} &= {{\beta }_{m}}{{\delta }_{mn}}. \label{param:CMEB}
\end{align}

$\mathbf{B}$ is the propagation constant matrix, with diagonal terms representing propagation constants. $\mathbf{K}^{+}$ is the intermodal coupling matrix for co-propagating modes,and K- the back-reflection coupling matrix for counter-propagating modes. $\mathbf{E}_{n}$ and $\mathbf{H}_{n}$ are the electric field vector and magnetic field vector of $n^{th}$ eigenmode.

By applying the chain rule, we can express the coupling element as the product of the geometric gradient and an intrinsic waveguide parameter, such as \ref{geo_param}, where $h$ means height of the waveguide and the $f$ means the frequency. One interesting fact is that this parameter has been widely used in the design of adiabatic devices as a kind of hybrid factor to guide how slowly the device structure should change\cite{chung_short_2017,chung_fast_2019,siriani_adiabatic_2021,tambasco_expanding_2023,wang_design_2023}.

\begin{equation}
\begin{split}
  K_{mn}^{+} &= 
  \frac{\frac{dw}{dy} \iint\limits_{xOz}{\left( 
  \frac{\partial \mathbf{E}_n}{\partial w} \times \mathbf{H}_m^\dagger 
  + \mathbf{E}_m^\dagger \times \frac{\partial \mathbf{H}_n}{\partial w} \right)} 
  \cdot d\mathbf{S}_{zx}}
  {\iint\limits_{xOz}{\left( 
  \mathbf{E}_m \times \mathbf{H}_m^\dagger 
  + \mathbf{E}_m^\dagger \times \mathbf{H}_m \right)} \cdot d\mathbf{S}_{zx}} \\
  &= \frac{dw}{dy} G_{mn}^{+}(w; h, f, \dots).
\end{split}
\label{geo_param}
\end{equation}

We can use the eigenmode solver to extract the intrinsic parameter. Based on the distribution of the extracted parameter, it is wise to choose conformal subcell or logarithmic equidistant subcell for eigenmode analysis. In parameter extraction, energy normalization and phase alignment can help to obtain an accurate and smooth parameter matrix.

\begin{equation}
\begin{split}
  j\frac{d}{dy}\begin{bmatrix}
    \varphi^{+} \\
    \varphi^{-}
  \end{bmatrix} &= 
  \left( 
  \begin{bmatrix}
    \mathbf{B} & 0 \\
    0 & -\mathbf{B}
  \end{bmatrix}
  - j\frac{dw}{dy}
  \begin{bmatrix}
    \mathbf{G}^{+} & \mathbf{G}^{-} \\
    \mathbf{G}^{-} & \mathbf{G}^{+}
  \end{bmatrix}
  \right)
  \begin{bmatrix}
    \varphi^{+} \\
    \varphi^{-}
  \end{bmatrix} \\
  &= 
  \mathbf{H}(y)
  \begin{bmatrix}
    \varphi^{+} \\
    \varphi^{-}
  \end{bmatrix}.
\end{split}
\label{eq:EPM}
\end{equation}

Once the intrinsic parameters are extracted, we can leverage the geometric gradient (e.g. $dw/dy$) to construct compact model and obtain the frequency domain transmission characteristics of the device by solving corresponding equations\ref{eq:EPM}. To ensure physical accuracy, the solution of the differential equations must preserve unitarity.

A relatively simple approach for solving such equations can be derived based on the one-order superadiabatic approximation\cite{berry_quantum_1987,guery-odelin_shortcuts_2019,martinez-garaot_fast_2015,garrido_generalized_1964}. Thus the time-dependent Hamiltonian can be quasi-diagonalized\ref{eq:SA_approx}. This approximation provides a computationally efficient method while maintaining the essential characteristics of the system's dynamics. The recurrence formula is as follows\ref{eq:recurrence}

\begin{equation}
i\frac{d{{\mathbf{U}}^{\dagger }}\varphi }{dy}=\left( \mathbf{D}-i{{\mathbf{U}}^{\dagger }}\frac{d\mathbf{U}}{dy} \right){{\mathbf{U}}^{\dagger }}\varphi \approx \mathbf{D}{{\mathbf{U}}^{\dagger }}\varphi \label{eq:SA_approx}
\end{equation}

\begin{equation}
    \varphi \left[ i+1 \right]=\mathbf{U}\left[ i+\frac{1}{2} \right]{{e}^{-i\mathbf{D}\left[ i+\frac{1}{2} \right]\Delta y}}\mathbf{U}{{\left[ i+\frac{1}{2} \right]}^{\dagger }}\varphi \left[ i \right]\label{eq:recurrence}
\end{equation}

Where $\mathbf{U}$ is the normlized eigenvector of $\mathbf{H}$, and $\mathbf{D}$ is the eigenvalue of $\mathbf{H}$. In this formula, we ignore the coupling term, which becomes small enough when the number of subcells increases.

This method not only accurately captures the continuous smooth geometric structure but also effectively handles the step-like effects in the geometry, like $dw/dy$ approaches the Dirac-delta function. The key approach is to replace the parameter $y$ in the equation with $w$, thereby transforming the infinity into zero and simplifying the calculation by neglecting the propagation constant matrix $\mathbf{B}$.

\begin{figure}[hb]
\centering
\includegraphics[width=0.9\textwidth]{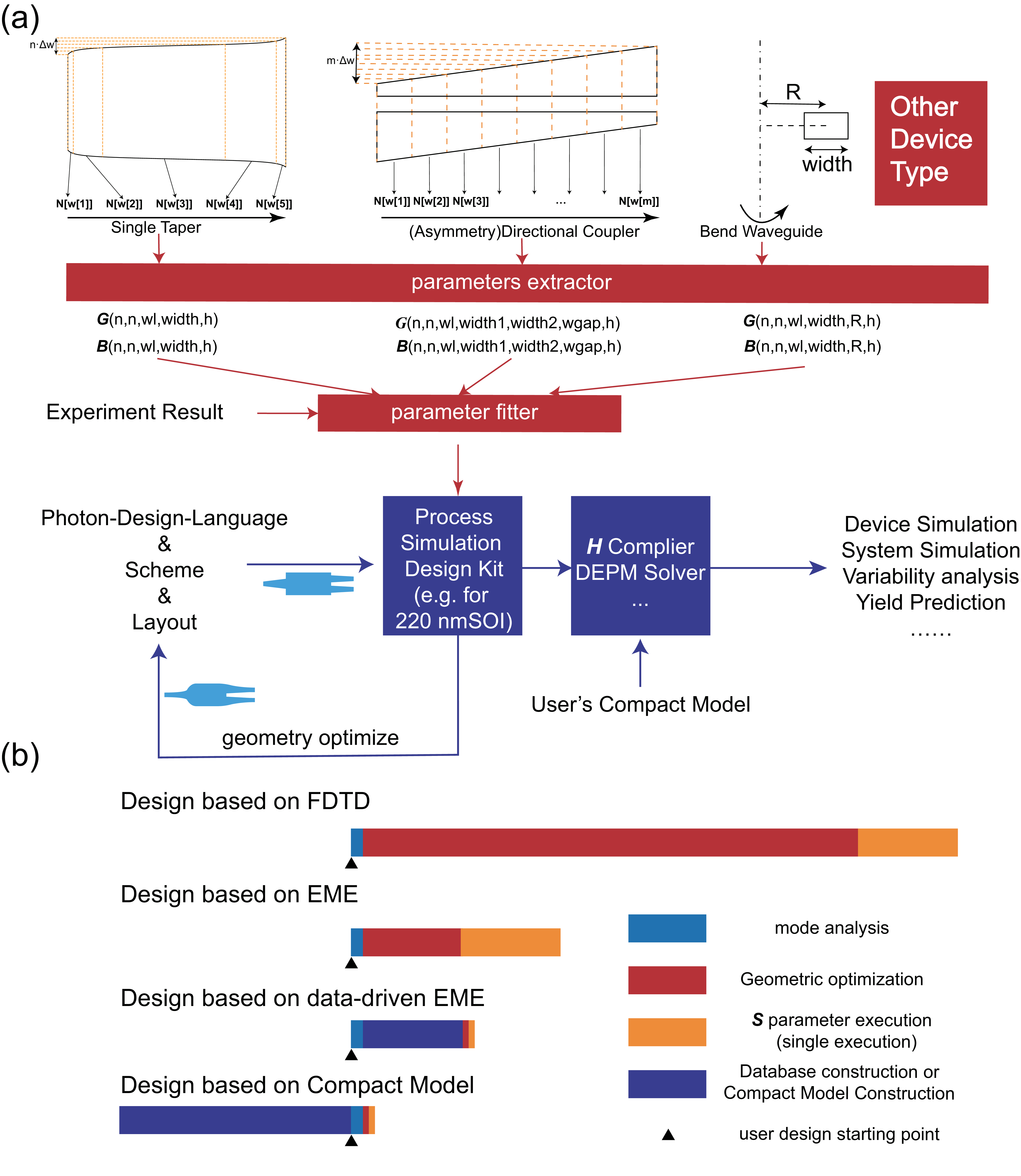} 
\caption{%
(a) Photon design automation workflow based on compact model of linear passive optical waveguide 
(b) Qualitative comparison of simulation time based on different tools.}
\label{fig_PDSK}
\end{figure}

Figure \ref{fig_PDSK}(a) shows a possible workflow for future photon design automation based on this compact model. Foundries or simulation software developers can extract and fit compact models using high-precision physical simulation tools such as FDTD and EME for specific process platforms, and fine-tune the compact models according to experimental results. When all is done, this set of data can be packaged as a Process Simulation Design Kit (PSDK), which allows users to describe their devices at multiple levels and rely on PSDK for multiple levels of device simulation, geometric optimization, system simulation, variability analysis, yield prediction, etc., to adapt to highly customized device operating characteristics.

Figure \ref{fig_PDSK}(b) shows qualitative time comparisons for device design, optimization, and parameter extraction based on different simulation tools. FDTD performs full wave analysis of the device and consumes a lot of time in geometric optimization. EME method has certain advantages in parameter scanning, and can efficiently evaluate the performance of directional couplers, adiabatic waveguides, multimode interferometers, etc., but it is still necessary to re-analyze the eigenmodes when the wavelength changes or errors are introduced. The data-driven EME algorithm records the scattering matrix, which may require more time, but reduces the marginal consumption of subsequent calculations. The compact model-based design proposed in this work may be able to improve efficiency in all aspects by using PSDK provided by foundries and EPDA tool suppliers in the future

\section{Case study and numerical experiment results}\label{sec4}

To verify the proposed compact model, a $TM_{0}\text{--}TE_{1}$ mode converter is selected as a typical case for theoretical validation. Figure 2 illustrates the workflow of the mode converter design, optimization, and tolerance analysis based on compact model extraction and the Data-driven Eigenmode Propagation Method.

\begin{figure}[hb]
\centering
\includegraphics[width=0.9\textwidth]{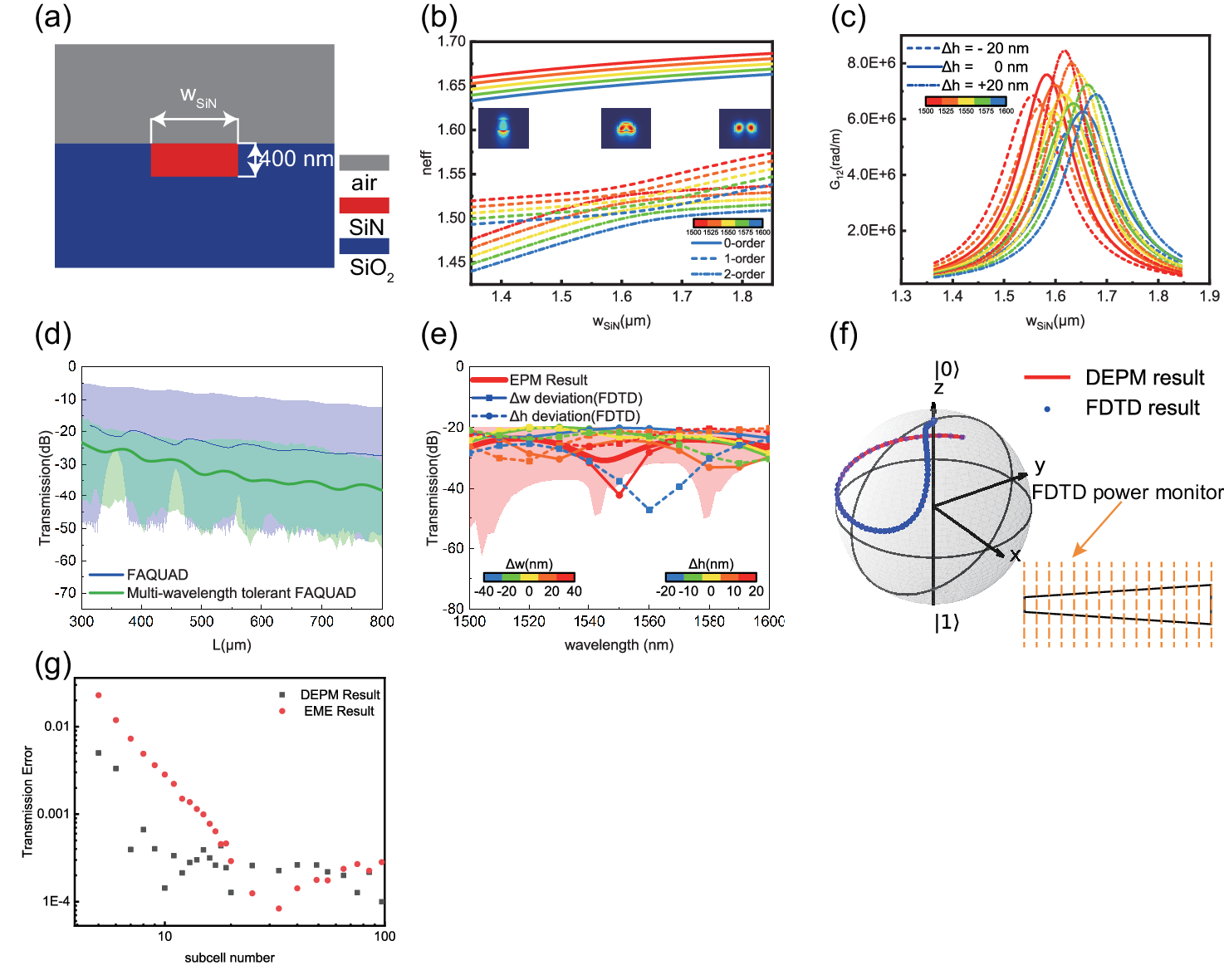} 
\caption{%
Design process of mode converter based on compact model (a) The cross section structure of the mode converter (b) the change of the effective refractive index of the mode with wavelength and width (c) the change of the non-diagonal element with wavelength and thickness (d) The length, wavelength scanning and tolerance analysis of the device were performed using DEPM (e) the correctness of the DEPM tolerance analysis was verified using FDTD (f) Comparison of Bloch team FDTD and DEPM calculation results (g) Comparison of algorithm convergence between DEPM and EME
The results, shown in Figures 2(b) and 2(c), reveal the presence of propagation}
\label{fig_ModeConverter}
\end{figure}

Figure \ref{fig_ModeConverter}(a) depicts the cross-sectional structure of the mode converter implemented on a 400 nm LP-SiN platform. To achieve effective coupling between orthogonal polarization modes, an air envelope is introduced to break the vertical symmetry of the waveguide structure. Using the Ansys Lumerical FDE solver, parameter sweeps were conducted on silicon nitride waveguides with various widths. A fixed and fine mesh was applied, with a 5 nm grid for the waveguide, a 10 nm grid for the evanescent decay region, and default PML(perfect matching layer) settings. After normalizing the eigenmodes by their Poynting vector within the non-PML region, the mode coupling coefficient matrix was calculated using Equation \ref{geo_param}. The effective refractive indices and coupling coefficients computed were stored as a lookup table. 

The results, shown in Figures \ref{fig_ModeConverter}(b) and (c), reveal the presence of propagation constants anti-crossings. Each nondiagonal element of the coupling coefficient matrix closely follows a Lorentz distribution, with an R-value exceeding 0.9997. Based on this, two curves were designed: one following the single-wavelength FAQUAD theory\cite{martinez-garaot_fast_2015} and another supplemented with multiwavelength tolerance considerations\cite{chung_shortcut_2022}. Using the DEPM method, comprehensive parameter sweeps were performed to account for variations in device length, waveguide width, and fabrication deviations. A Monte Carlo algorithm was employed to resample the parameter sweep results, obtaining the statistical relationship between crosstalk and device length, as shown in Figure \ref{fig_ModeConverter}(d). The solid line represents the predicted length-crosstalk curve at 1550 nm without fabrication deviations, while the shaded region indicates the 99.7 \% confidence interval that includes manufacturing variability and wavelength bias.

Ultimately, a device length of 395 $\mu$m was selected as the shortest length satisfying the -20 dB crosstalk requirement.  High-precision FDTD simulations were conducted for verification. At 11 wavelength points and 9 fabrication deviation scenarios, the FDTD simulation results perfectly matched the DEPM predictions. Statistically, all predicted crosstalk values were below -19.6 dB.

To further investigate the accuracy of the method, the mode converter length was set to 93 $\mu m$, with linear geometric transitions to deviate from adiabaticity. High-precision FDTD simulations were performed using a 5 nm mesh, with 101 monitors tracking mode evolution. In this mode converter, the optical waveguide can be regarded as a two-level system composed of the $1^{st}$ and $2^{nd}$ eigenmodes, and its evolution can be described by Bloch spheres. Figures \ref{fig_ModeConverter}(f) compared FDTD and DEPM results on the Bloch sphere, which demonstrated a high degree of consistency.

Finally, to validate the efficiency of parameter prediction in reducing sampling points, Lorentzian fits were applied to the parameters obtained from different subcell numbers.  DEPM algorithms were executed based on these fits, and convergence diagrams were generated using FDTD as the ground truth. The results shows in Figures \ref{fig_ModeConverter}(g), confirming that the proposed approach significantly reduces sampling requirements while achieving higher parameter extraction efficiency than data-driven EME methods.

In this case study, the construction of the compact model took approximately 300 minutes, covering a wavelength range from 1500 nm to 1600 nm and accounting for a thickness variation of 20 nm. Subsequently, we used interpolation to supplement the data and performed parameter sweep scans across 51 different lengths, 5 variations in width, 5 variations in waveguide thickness, and 11 wavelengths using proposed DEPM method. Using the FDTD method with GPU acceleration, this process would require more than 10,000 hours; with the EME method, around 275 hours. In contrast, DEPM completed the task in less than 3 hours. If applied to more complex optimization processes and improved differential equation solving algorithms, our approach shows greater potential for time efficiency.

\section{Efficient compact model parameter extraction}\label{sec3}
Efficient parameter extraction is crucial for our compact model based on unitary evolution. By leveraging the physical properties of waveguide devices and selecting appropriate physical models as references, the parameter extraction process can be significantly improved through fitting methods. This ensures a low noise parameter distribution and enhances numerical stability when using recurrence formula \ref{geo_param}.

Key parameters in the proposed compact model include the effective refractive index and the structural matrix $\mathbf{G}$. While the fitting for $n_{eff}$ has been extensively studied, the behavior of $\mathbf{G}$ remains underexplored. Our findings indicate that, for most devices, the dependence of $\mathbf{G}$ on the waveguide width $w$ can be described using power-law or Lorentzian-like distributions, which can be understood as resonant transitions and nonlinear transitions in multi-body systems.

\begin{figure}[ht]
\centering
\includegraphics[width=0.95\textwidth]{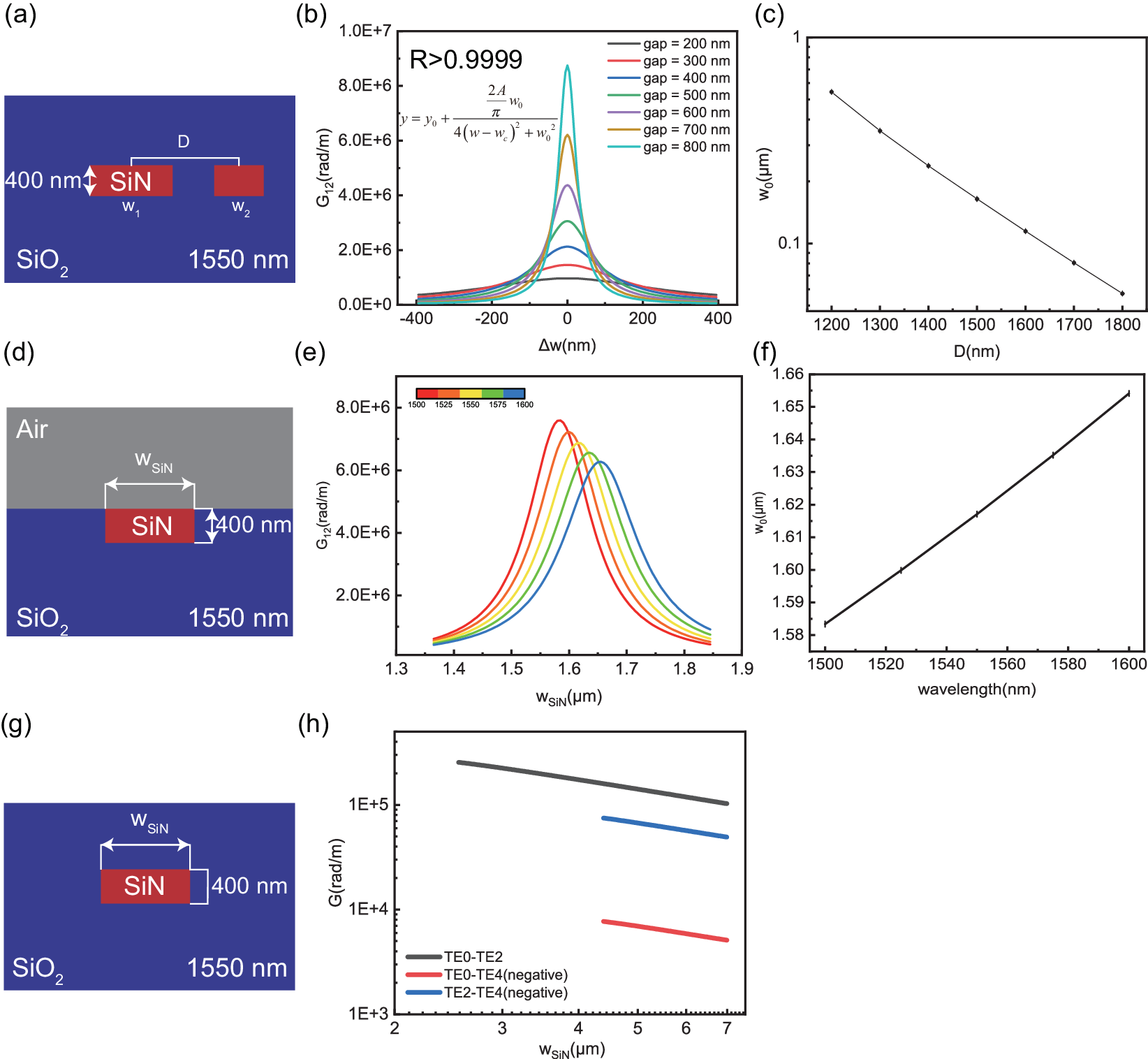} 
\caption{%
The distribution of $\mathbf{G}$ parameters (a) The cross-sectional structure of the coupler device (b) The variation of the non-diagonal element of parameter $\mathbf{G}$ with $\Delta w$ difference under different $w_{gap}$ (c) The change of the FWHM on $\mathbf{G}$ with $w_{gap}$, with ordinate logarithmic scale (d) Section structure of mode converter on 400 nm SiN , with air upper clad (e) The variation of the non-diagonal element of parameter $\mathbf{G}$ with $\Delta w$ difference under different wavelength (f) Variation of parameter w0 with wavelength (g) Section structure of multimode waveguide on 400 nm SiN (h) The variation of the non-diagonal element of parameter $\mathbf{G}$ with $\Delta w$ 
The results, shown in Figures 2(b) and 2(c), reveal the presence of propagation}
\label{fig_fit}
\end{figure}

Including the intrinsic parameters with Lorentzian distributions already shown in Figure \ref{fig_ModeConverter}(c), we observed that the off-diagonal elements of the structural matrix $\mathbf{G}$, related to coupler-type structures such as $TE_{0}$ or $TM_{0}$ even-odd supermodes in adiabatic couplers and $TE_{0}\text{--}TE_{1}$ modes in asymmetric directional couplers, also exhibit Lorentzian distributions dependent on the waveguide width $w$. Other fitting parameters are specific to the waveguide, such as the center-to-center distance $D$, waveguide thickness, and operating wavelength. Figure \ref{fig_fit} presents some of our computational results, all calculated on 1550 nm in default. 

Figure \ref{fig_fit}(a) shows a dual-core structure commonly found in adiabatic couplers\cite{cook_tapered_1955,yin_ultra-broadband_2017}, where the width sum remains constant, but the width difference ($w_{1}-w_{2}$) is varied with y. On a 400 nm Silicon Nitride platform, we analyzed the $0^{th}$-order and $1^{st}$-order supermodes, finding that the off-diagonal elements of the structural matrix G also follow a Lorentzian-like distribution, as shown in Fig. 3(b), with an R value exceeding 0.9999. Figure \ref{fig_fit}(c) demonstrates that FWHM decays exponentially as the coupling spacing $D$ increases. This trend can be explained using the complex coupled-mode equations from a non-supermode perspective, as mentioned in Appendix \ref{secA1}, where the coupling coefficient decreases exponentially with increasing D\cite{guo_compact_2018,guo_silicon_2017}.

Figure \ref{fig_fit}(d) illustrates the cross-sectional structure of a silicon nitride waveguide, where vertical symmetry is broken by an air cladding. Figure \ref{fig_fit}(e) displays a portion of the curves selected from Figure \ref{fig_ModeConverter}(c). After fitting, we found that the results strongly adhere to a Lorentzian distribution, with an $R$ value exceeding 0.9999. Furthermore, when the waveguide thickness remains constant, its center width $w$ exhibits a linear relationship with the wavelength, as shown in Figure \ref{fig_fit}(f).

Figure \ref{fig_fit}(g) depicts the cross-sectional structure of a multimode 400 nm silicon nitride waveguide, commonly found in MMI or Y-branch structures. Due to the rapid growth of structural parameters, coupling may occur between modes that satisfy even or odd symmetry, such as from the $TE_{0}$ mode to the $TE_{2}$ or $TE_{4}$ modes. This result is illustrated in Figure \ref{fig_fit}(h), where the curves appear linear in a double-logarithmic coordinate system, following a power-law distribution with an R value exceeding 0.99999.

Through these numerical and fitting results, we conclude that studying the distribution patterns of the off-diagonal elements in the structural matrix $\mathbf{G}$ in advance, summarizing applicable models, and subsequently conducting parameter extraction of compact model based on these models not only accurately describe the unitary evolution characteristics of various waveguide device elements but also significantly reduce instability caused by parameter fluctuations during subsequent numerical simulations. This approach provides a fast and reliable modeling technique for the construction of proposed PDSK.

These findings further inspire us to optimize the sampling process in parameter extraction. Instead of performing exhaustive sampling across all grid points, grid sampling can be customized based on the expected distribution characteristics. For example, for Lorentzian distributions, conformal grid sampling may be the optimal choice. In cases where the off-diagonal elements follow a power-law distribution, logarithmically equidistant grids can be used to minimize numerical errors, as suggested by the guidelines in reference\cite{oktay_computationally_2024}.

However, certain limitations of our current results should be noted. For instance, when transitioning from bound modes to radiative continuum modes, a unified model for describing the off-diagonal elements of the structural matrix G has yet to be identified. In some cases, the distribution follows a Hill function, while in others, it aligns with a Boltzmann model. This inconsistency may be related to the settings of PML in eigenmode calculations and warrants further investigation to understand its influence.

\section{Conclusion}\label{sec5}
Compared with commonly used FDTD and EME, DEPM connects multiple fields such as underlying physics, data-driven, efficient simulation, and tolerance analysis from the perspective of unitary evolution, so that it can often achieve simulation accuracy comparable to 3D-FDTD with fewer grid points or more simplified subunits.In applications that do not involve numerical apertures and mode numbers such as diffraction gratings or flat layers, DEPM can be an efficient tool to bridge the gap between personalized design, manufacturing-oriented design, and photonic design automation. It was not created to replace FDTD, EME, BPM and other algorithms, but to provide a new perspective and choice for industry and academia. For extremely complex three-dimensional nanostructures, strong nonlinear or high-loss metal waveguides, traditional full-wave simulations (such as FDTD) are still irreplaceable accurate benchmarks.

In the future, leveraging high-precision simulations and experimental validations for parameter extraction in the proposed compact models, EPDA providers may introduce the described Process Simulation Design Kits (PSDKs) as a complement to traditional PDKs. These kits could provide a unified framework for photonic design automation workflows, encompassing device design and system optimization. Designers could use PSDKs to rapidly optimize passive components for specific project requirements without the need to directly solve computationally intensive Maxwell equations.

The proposed method holds significant potential to accelerate design and iteration cycles while maintaining the high accuracy required for advanced photonic systems. As photonic technologies continue to evolve, this approach could pave the way for a new generation of efficient, adaptive, and scalable design tools, addressing the growing complexity and customization demands of integrated photonics.

\backmatter

\section{Supplementary information}
The time evaluations were conducted on a system equipped with an Intel i9-10850K processor, 64 GB DDR4 memory, and a NVIDIA Quadro RTX 5000 professional graphics processor.

\section*{Declarations}

\subsection*{Funding}
Not applicable. 

\subsection*{Conflict of interest/Competing interests}
The author declares no conflict of interest.

\subsection*{Ethics approval and consent to participate}
Not applicable.

\subsection*{Consent for publication}
Not applicable.

\subsection*{Data availability}
The datasets generated and analyzed during the current study are available from the corresponding author upon reasonable request.

\subsection*{Materials availability}
Not applicable.

\subsection*{Code availability}
The code used for this study is available from the corresponding author upon reasonable request.

\subsection*{Author contribution}
Zijian Zhang conducted all aspects of this research, including conceptualization, methodology development, data analysis, and manuscript preparation.

\bigskip

\begin{appendices}

\section{Physical interpretation of Lorentz distribution of nondiagonal elements of G matrix with coupler structure}\label{secA1}

Based on the non-orthogonal coupled mode theory we can quantify the coupling behavior of two local modes in two waveguides. The formula is \ref{eq:non_orth_CME}
\begin{equation}
j\left[ \begin{matrix}
   {{N}_{00}} & {{N}_{01}}  \\
   {{N}_{10}} & {{N}_{11}}  \\
\end{matrix} \right]\frac{d}{dy}\left[ \begin{matrix}
   {{A}_{0}}  \\
   {{A}_{1}}  \\
\end{matrix} \right]=\left[ \begin{matrix}
   {{\chi }_{00}}+{{\beta }_{0}} & {{\chi }_{01}}  \\
   {{\chi }_{10}} & {{\chi }_{11}}+{{\beta }_{1}}  \\
\end{matrix} \right]\left[ \begin{matrix}
   {{A}_{0}}  \\
   {{A}_{1}}  \\
\end{matrix} \right]+\frac{dw}{dy}\left[ \begin{matrix}
   {{S}_{00}} & {{S}_{01}}  \\
   {{S}_{10}} & {{S}_{11}}  \\
\end{matrix} \right]\left[ \begin{matrix}
   {{A}_{0}}  \\
   {{A}_{1}}  \\
\end{matrix} \right]\label{eq:non_orth_CME}    
\end{equation}

where

\begin{align}
  & {{N}_{ij}}=\iint\limits_{zOx}{\hat{y}\cdot \left( \mathbf{\tilde{E}}_{j}^{*}\times {{{\mathbf{\tilde{H}}}}_{i}}+{{{\mathbf{\tilde{E}}}}_{i}}\times \mathbf{\tilde{H}}_{j}^{*} \right)dzdx}\label{param:N} \\ 
 & {{\chi }_{ij}}=\omega {{\varepsilon }_{0}}\iint\limits_{zOx}{\left( {{\varepsilon }_{r}}-{{\varepsilon }_{rj}} \right)\mathbf{\tilde{E}}_{i}^{*}\cdot {{{\mathbf{\tilde{E}}}}_{j}}dzdx}\label{param:chi} \\ 
 & {{S}_{ij}}=\iint\limits_{zOx}{\left( E_{ix}^{*}\frac{\partial {{H}_{jz}}}{\partial w}-E_{iz}^{*}\frac{\partial {{H}_{jx}}}{\partial w}+H_{ix}^{*}\frac{\partial {{E}_{jz}}}{\partial w}-H_{iz}^{*}\frac{\partial {{E}_{jx}}}{\partial w} \right)dzdx}\label{param:S}
\end{align}

Under the weak coupling condition, every matrix element of the matrix S is smaller. This equation can be further simplified as \ref{eq:sim_CME}
\begin{equation}
 j\left[ \begin{matrix}
   {{N}_{00}} & {{N}_{01}}  \\
   {{N}_{10}} & {{N}_{11}}  \\
\end{matrix} \right]\frac{d}{dy}\left[ \begin{matrix}
   {{A}_{0}}  \\
   {{A}_{1}}  \\
\end{matrix} \right]=\left[ \begin{matrix}
   {{\chi }_{00}}+{{\beta }_{0}} & {{\chi }_{01}}  \\
   {{\chi }_{10}} & {{\chi }_{11}}+{{\beta }_{1}}  \\
\end{matrix} \right]\left[ \begin{matrix}
   {{A}_{0}}  \\
   {{A}_{1}}  \\
\end{matrix} \right]\label{eq:sim_CME}   
\end{equation}
After a series of transformations, the equation is finally reduced to the standard two-level coupled mode equation\ref{eq:twolevel}
\begin{equation}
    j\frac{d}{dy}\left[ \begin{matrix}
   {{A}_{0}}  \\
   {{A}_{1}}  \\
\end{matrix} \right]=\left[ \begin{matrix}
   \Delta  & \Omega   \\
   \Omega  & -\Delta   \\
\end{matrix} \right]\left[ \begin{matrix}
   {{A}_{0}}  \\
   {{A}_{1}}  \\
\end{matrix} \right]\label{eq:twolevel}
\end{equation}
The unitary transformation of the state vector is performed according to the eigenvector of the instantaneous Hamiltonian, and the formula is finally obtained\ref{eq:sim_SA}
\begin{equation}
    j\frac{d{{\mathbf{U}}^{\dagger }}\mathbf{A}}{dy}=\left[ \begin{matrix}
   \sqrt{{{\Delta }^{2}}+{{\Omega }^{2}}} & -\frac{1}{2}j\dot{\theta }  \\
   \frac{1}{2}j\dot{\theta } & -\sqrt{{{\Delta }^{2}}+{{\Omega }^{2}}}  \\
\end{matrix} \right]{{\mathbf{U}}^{\dagger }}\mathbf{A}\label{eq:sim_SA}
\end{equation}
where
\begin{align}
  & \theta =\arctan \frac{\Omega }{\Delta } \\ 
 & \mathbf{U}=\left[ \begin{matrix}
   \cos \frac{\theta }{2} & \sin \frac{\theta }{2}  \\
   \sin \frac{\theta }{2} & -\cos \frac{\theta }{2}  \\
\end{matrix} \right]  
\end{align}
When the $\Delta w$ does not vary very much, it can be found that the propagation constant difference $\Delta$ follows a linear distribution with a slope of $k$, so that the non-diagonal elements can be simply derived\ref{eq:Lorentz_theory}
\begin{equation}
\begin{split}
    G_{12} & = -j \frac{dy}{dw} \left( -\frac{1}{2} j \dot{\theta} \right) \\
    & = -\frac{1}{2} \frac{dy}{dw} \frac{\Delta \frac{d\Omega}{dy} - \Omega \frac{d\Delta}{dy}}{\Delta^2 + \Omega^2} \\
    & \approx \frac{1}{2} \frac{k}{\Omega} \frac{1}{\left( \frac{k}{\Omega}w \right)^2 + 1}.
\end{split}\label{eq:Lorentz_theory}
\end{equation}
In this way, we can naturally derive the Lorentz distribution followed by the nondiagonal elements of the structural parameters. If we can find a suitable method to perform similar non-orthogonal decomposition of $TM_{0}-TE_{1}$ hybrid modes in waveguides with broken vertical symmetry, we may be able to theoretically explain the non-diagonal elements of the structural parameter G of a series of devices, including $TM_{0}-TE_{1}$ mode converters and $TM_{1}-TE_{2}$ mode converters.
\end{appendices}


\bibliography{sn-bibliography}

\end{document}